\documentclass[fleqn,usenatbib]{mnras}

\usepackage{newtxtext,newtxmath}

\usepackage[T1]{fontenc}

\DeclareRobustCommand{\VAN}[3]{#2}
\let\VANthebibliography\thebibliography
\def\thebibliography{\DeclareRobustCommand{\VAN}[3]{##3}\VANthebibliography}

\usepackage{graphicx}	
\usepackage{amsmath}	

\title[No X-ray Emission from SN 2018fhw]{Chandra Fails to Detect X-ray Emission from Type Ia SN 2018fhw/ASASSN-18tb}

\author[V. V. Dwarkadas]{
Vikram V. Dwarkadas$^{1}$\thanks{E-mail: vikram@astro.uchicago.edu}
\\
$^{1}$Department of Astronomy and Astrophysics, University of Chicago, 5640 S Ellis Ave., ERC 569, Chicago, IL 60637
}

\date{Accepted XXX. Received YYY; in original form ZZZ}

\pubyear{2022}

\begin{document}
\label{firstpage}
\pagerange{\pageref{firstpage}--\pageref{lastpage}}
\maketitle

\begin{abstract}
We report on Chandra X-ray observations of ASASSN-18tb/SN 2018fhw, a low luminosity Type Ia supernova that showed a H line in its optical spectrum. No X-ray emission was detected at the location of the SN. Upper limits to the luminosity of up to 3 $\times 10^{39}$ erg s$^{-1}$ are calculated, depending on the assumed spectral model, temperature and column density. These are compared to two Type Ia-CSM SNe,  SN 2005gj and SN 2002ic, that have been observed with Chandra in the past. The upper limits are lower than the X-ray luminosity found for the Type Ia-CSM SN 2012ca, the only Type Ia SN to have been detected in X-rays. Consideration of various scenarios for the H$\alpha$ line suggests that the density of the surrounding medium at the time of H$\alpha$ line detection could have been as high as 10$^8$ cm$^{-3}$, but must have decreased below 5 $\times\, 10^6$ cm$^{-3}$ at the time of X-ray observation. Continual X-ray observations of SNe which show a H line in their spectrum are necessary in order to establish Type Ia SNe as an X-ray emitting class.
\end{abstract}




\section{Introduction}

Supernovae (SNe) are primarily divided into two types. SNe of Type I do not contain H in their spectra, whereas those of Type II show H in their early spectra. Each type is further subdivided into several subtypes and subclasses, based on their early time spectra or lightcurves.  Type Ia SNe (Ia's for short) show a distinct Si line in their spectra. They have been used to measure the expansion and acceleration of the universe, a long-time endeavor that was awarded a Nobel prize in physics in 2011. Given their significance, it is concerning that we do not yet know the nature of their progenitor systems.

It is generally accepted that the progenitor of a SN Ia must be a white dwarf. In order for the white dwarf to explode, its mass must increase until it exceeds  the Chandrasekhar mass, although some calculations suggest that white dwarfs below the Chandrashekar mass can also explode in certain circumstances \citep{ruiter20}. It is likely that the white dwarf  must have accreted mass transferred from a companion in a binary system. The nature of the companion is hotly debated \citep{pilar14}, with evidence existing for both double-degenerate and single degenerate systems \citep{maoz14}. In the former case, the companion is another white dwarf \citep{scalzoetal10, silvermanetal11, nugentetal11, bloometal12, brownetal12}, while in the latter case it is a main sequence or evolved star \citep{hamuyetal03, dengetal04}. If the latter, the SN can accrete material from the donor companion star, raising its mass over the Chandrasekhar limit, leading to an explosion. Difficulties arise in either scenario, and it is likely that multiple channels exist to form Type Ia SNe \citep{lm18}. \citet{soker19} has argued that  Type Ia SNe must be spherical given the low polarisation measurements, and that the detonation to deflagration explosion of a Chandrashekhar mass white dwarf can result in a global spherical structure that is compatible with the observations. In recent years however many objects have been found which have a spectrum resembling a Type Ia, but deviate in some manner from the parameters typical of Ia's \citep{taubenberger17}, and consequently a large number of models exist to explain them \citep{rs18,tn19}.

In the scenario where the companion is a main-sequence or evolved star, there exists the possibility of H-rich material around the progenitor, arising from mass loss by the companion star. The SN shock wave expanding outwards can interact with this H-rich medium, thus resulting in the presence of H in the optical spectrum. In an alternate scenario, the H may be swept-up from the companion star. Either way, these SNe would show H lines in their optical spectrum. Some Type Ia SNe have been observed that exhibit narrow hydrogen lines superimposed on a SN Ia-like spectrum \citep{hamuyetal03,dengetal04}. The narrow line width implies a velocity lower than that of the typical Type Ia SN shock velocity. A possible explanation is that it arises from the SN shock expansion into a dense surrounding medium. These SNe ($<  20$ in number) have been grouped together under the subclass of Type Ia-CSM \citep{silvermanetal13}. They have a distinct H line, with H$\alpha$ luminosities of order 10$^{40}$ to 10$^{41}$ erg s$^{-1}$, Balmer decrements (ratio of H$\alpha$ to H$\beta$ intensity) higher than the nominal ratio of 3, and larger absolute magnitudes compared to normal Type Ia's.

The presence of a H line in the optical spectrum of a Type Ia may indicate some level of interaction with a dense surrounding medium, not generally expected around normal Type Ia SNe. Signposts of circumstellar interaction in SNe include  X-ray and radio emission. Over 60 SNe have been detected till now in X-rays \citep{dg12, vvd14, drrb16, rd17, bocheneketal18}, and the number is growing quickly. Many others have upper limits calculated. All but one have been of the core-collapse variety. The singular exception is SN 2012ca,  which represents the first and only detection of X-ray emission from a Type Ia SN \citep{bocheneketal18}. SN 2012ca, which belongs to the Type Ia-CSM category, was observed twice with the Chandra X-ray telescope. The first X-ray observation of SN 2012ca, at 554 days past explosion, resulted in 33 counts and a solid detection. The second at 745 days yielded 10 counts. The measured Balmer decrement  ranged from 3-20. The most conservative estimate of the surrounding density yielded a value of at least 10$^6$ particles cm$^{-3}$. The authors preferred a two-component medium, with the X-ray emitting material having a much higher density, around 10$^8$ cm$^{-3}$. In either case, it was clear that in order to reach the observed X-ray intensity, the SN must have been surrounded by a dense medium with which the shock interacted to produce the observed X-rays.

The detection of a single Type Ia-CSM in X-rays was in itself highly significant. However, in order to understand the physics of these objects, decipher their progenitors, and investigate their environment, it is important to detect more SNe having similar characteristics. There have been arguments against SN 2012ca being a Type Ia SN \citep{inserraetal14}, although they were contradicted by  \citet{foxetal15}. Nonetheless, it is true that the entire class of Ia-CSM SNe has properties that stand out from the majority of Type Ia SNe. In order to conclusively establish a new class of X-ray emitting objects, it is imperative that we find, and detect with higher significance, classical Type 1a SNe in X-rays. While evidence has been mounting that Type Ia's arise from double degenerate systems, the detection of a Type 1a SN in X-rays would likely indicate the presence of a dense ambient medium, which could be stripped off the companion, or arise from a previous mass-loss episode of the companion star. In either case this would signify a single degenerate progenitor. Past searches for X-ray emission from Type Ia SNe \citep{marguttietal14, sandetal21} have generally failed to detect any sources besides SN 2012ca, even from SNe that showed a H-line in the optical spectrum \citep{hughesetal07}. 

In this paper we report on recent Chandra observations of  the SN ASASSN-18tb/SN 2018fhw (hereafter referred to as SN 2018fhw), which also showed a H-line in its optical spectrum. In \S \ref{sec:2018fhw} we review optical observations of SN 2018fhw, as well as those of similar Type Ia SNe that have recently been detected which showed a H line in their spectrum. In \S \ref{sec:obs} we provide details of the X-ray observation, and compute upper limits to the X-ray flux and intrinsic luminosity of SN 2018fhw under various assumptions. In \S \ref{sec:comp} we compare the upper limits of SN 2018fhw to those of Type Ia-CSMs that have been observed in the past. \S \ref{sec:disc} estimates the density of the medium around the SN at the time that the H$\alpha$ line was detected, and the resultant X-ray luminosity, and then estimates parameters at the time of the Chandra observation. This is used to calculate the density of the medium at the X-ray epoch. Finally \S \ref{sec:results} summarizes our work and compares to that at other wavelengths.

\section{SN 2018fhw}
\label{sec:2018fhw}
SN 2018fhw was discovered \citep{brimacombeetal18} by the All-Sky Automated Survey for SNe \citep[ASAS-SN]{shappeeetal14}. Classification was based on a SALT spectrum \citep{eweisetal18}. A low dispersion spectrum obtained with LDSS3 on the Magellan Clay telescope at 139 days past explosion showed a distinct narrow H$\alpha$ line with a FWHM of 1085 km s$^{-1}$ \citep[][hereafter K19]{kollmeieretal19}. 

Using a redshift of 0.017, and a distance of 70.7 Mpc, K19 estimated a line flux of 2.2 $\pm\, 0.2\, \times \,10^{38}$ erg s$^{-1}$.  No obvious emission was observed at the position of H$\beta$, leading to an estimate of a Balmer decrement of at least 6, but potentially as large as 14. 

 \citet[][hereafter V19]{vallelyetal19} studied SN 2018fhw using the TESS and SALT telescopes. They find indications for an H$\alpha$ line starting on day 37, and can measure the line at least to day 148. They calculate a FWHM $\approx$ 1390 $\pm$ 200 km s$^{-1}$ up to 148d.  For spectra before about 100d they place a lower limit of 2 on the Balmer decrement, whereas for later spectra they place a limit of 5, consistent with the lower limit of 6 by K19.  Intriguingly, the H$\alpha$ flux in SN 2018fhw is found to be nearly constant by V19, with large error bars. The constant value of $\approx$ 2. $\times 10^{38}$ erg s$^{-1}$ is close to the value of 2.2 $\times 10^{38}$ erg s$^{-1}$ given for the flux by K19. The constant flux is interpreted by V19 as evidence that the H$\alpha$ arises from circumstellar interaction rather than being swept-up material from a non-degenerate companion. If the latter, the H$\alpha$ flux would be expected to follow the Fe III line flux, which it does not, as the Fe III flux is decreasing in time. It would also be expected to follow the SN bolometric luminosity, which it also does not.

Although SN 2018fhw had an optical spectrum that showed the presence of H$\alpha$, in many other respects it was quite different from the 1a-CSMs. It was a sub-luminous SN Ia, compared to the Ia-CSMs which are super-luminous Ia's. The Ia-CSMs are preferentially found in star-forming galaxies, whereas SN 2018fhw occurred in an early-type galaxy comprising of old stellar populations. Ia-CSMs generally show H$\alpha$ luminosities about 2 orders of magnitude higher than that found in SN 2018fhw. It is not clear if SN 2018fhw forms a low-luminosity member of the Ia-CSM class, or a different SN class. On the one hand, the observed H$\alpha$ line suggests some similarity in physical characteristics. On the other, there is no question that SN 2018fhw is a definite Type Ia SN, whereas doubts about the Ia origin of the Ia-CSMs still persist. 

In the past few years, 2 more low-luminosity Type Ia SNe have been discovered that showed a H line in their optical spectrum. Their classification as Type Ia's is equally secure.  SN 2016jae \citep{eliasrosaetal21} showed the presence of an H$\alpha$ line in two spectra 84 and 142 days after peak. The H$\alpha$ FWHM was 650 km s$^{-1}$ at the first epoch, and $\sim$ 1000 km s$^{-1}$ at the second. The line luminosity was 3 $\pm\, 0.8 \times 10^{38}$ erg s$^{-1}$ at the first epoch, and decreased by about a factor of two between the epochs; alternately, given the error bars it is possible that the luminosity may have been more or less constant. SN 2018cqj/ATLAS18qtd \citep{prietoetal20} also clearly showed the presence of an H$\alpha$ line at 193 and 307 days after peak. The line was resolved at the first epoch to have a FWHM of 1200 km s$^{-1}$ and a luminosity of $\approx 3.8 \times 10^{37}$ erg s$^{-1}$, with the luminosity decreasing by almost an order of magnitude at the second epoch.  These SNe are similar to SN 2018fhw in their low optical luminosity compared to traditional Type Ia SNe, and the luminosity and FWHM of the H$\alpha$ line. It is not clear if together these SNe form a continuum, or an entirely separate category from the Ia-CSMs. K19 suggested that up to 10\% of sub-luminous fast-declining SNe could show evidence of H$\alpha$ at nebular phases.  \citet{sandetal19} found no evidence for an H$\alpha$ line in a small sample of fast-declining Type Ia's, and suggested that they were rare phenomena. However, the subsequent discovery of two more subluminous Type Ia's with an H$\alpha$ line indicates that they may not be so rare. Of these three, only SN 2018fhw has been observed till now in X-rays.  An over-luminous Type Ia, SN 2015cp, had narrow H$\alpha$ detected nearly 700 days after explosion \citep{grahametal19}, much later than the ones above. The interaction had considerably decreased by about 800 days.

\section{X-ray Observations}
\label{sec:obs}
SN 2018fhw was observed with the ACIS-S detector on Chandra on July 27 2021 (ObsID 22445, PI: Dwarkadas). Given the discovery date of Aug 21 2018, the observation happened 1071 days after discovery, or about 1053 days in the rest frame of the SN. The Chandra TAC had approved the observation in 2019, for Cycle 21, but given the high ecliptic latitude, it was only observed during Cycle 22. The total exposure was 49.41 ks. The co-ordinates for SN 2018fhw are listed on Simbad as RA: 04 18 06.266  DEC: -63 36 54.25. However subsequent papers by the ASASSN group, as well as on the Transient Name Server (TNS), have amended the co-ordinates to 04:18:06.200 -63:36:56.41, about 2" away. We use the latter value throughout this paper, although our results are not sensitive to the 2" shift.

\begin{figure}
	\includegraphics[width=\columnwidth]{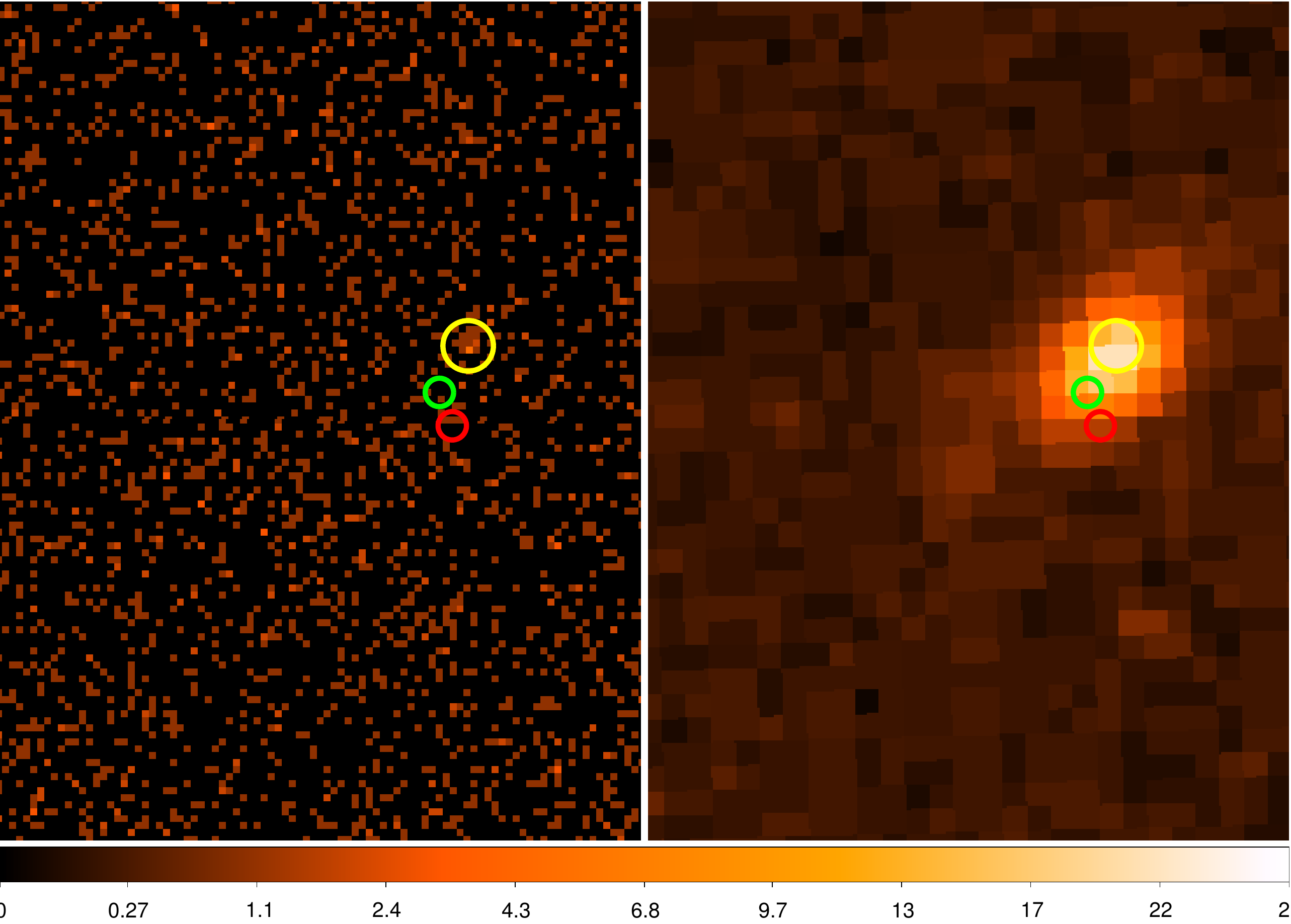}
    \caption{Left panel is the ACIS-S image. Right panel is the SAO DSS image matched to the exact same region. The Simbad position is marked with a green circle, the Transient Name Server one with a red circle. No X-ray emission is detected at either of these positions. The yellow circle encircles a nearby region ($\approx$ 5" away) where X-ray emission is detected. From the right panel, the yellow circle appears to correspond to the host galaxy, with the X-ray emission arising from near the center of the host galaxy.}
    \label{fig:18fhw}
\end{figure}

The left panel of Figure~\ref{fig:18fhw} shows the ACIS-S image of the region containing SN 2018fhw. The Simbad position is marked with a green circle, and the TNS one with a red circle. At neither of these positions is any X-ray emission detected. The yellow circle shows a nearby region ($\approx$ 5" away) where X-ray emission is detected. For comparison we show the SAO DSS image to the right, matched to the same area, and with the same regions overlaid. The yellow circle can be seen to correspond to the host galaxy. The X-ray emission appears to arise from near the center of the host galaxy. The red circle corresponding to the SN lies on the outskirts of the host galaxy.

A 1" region (as shown) includes 4 counts. However none of those counts lie in the 0.5-8 keV region. For the background region we choose an annulus stretching from 1" to 4" around the SN, keeping it clear of the X-ray emission from the host galaxy. We find 10 counts in this region between 0.5 and 8 keV. This would suggest 0.67 background counts in the source region. Inspecting the backgrounds at different locations and averaging, the number of counts in the background region are found to be within 15\% of the background value we find, therefore we consider the value to be robust. We use the Bayesian method of \citet{kbn91} to determine the maximum number of counts in the region, and derive 1.14/ 3.1/ 5.8 counts with a  68/ 95.4/ 99.7\% (1/ 2/ 3-$\sigma$) probability. In 49.41 ks this gives a maximum count rate of 2.3/ 6.3/ 11.74 $\times 10^{-5}$ counts s$^{-1}$, which is used to compute the upper limit on the flux. For additional confirmation, we also used the Chandra CIAO routine {\em srcflux} to calculate the flux within 0.5-8 keV. This routine returned an upper limit to the count rate of 6.23 $\times 10^{-5}$ counts s$^{-1}$, and a maximum flux of 8.48 $\times 10^{-16}$ erg s$^{-1}$ cm$^{-2}$ (2$\sigma$ values)  for a power-law model with an index of 2,  consistent with the values determined using the method of \citet{kbn91}. In the case of 3$\sigma$ values, the {\em srcflx} routine returns a count rate of 1.18 $\times 10^{-4}$ cts s$^{-1}$ and an unabsorbed flux in the 0.5-8 keV range of 1.6 $\times 10^{-15}$ erg s$^{-1}$ cm$^{-2}$, again consistent with our calculated values in Table \ref{tab:flux18fhw}. Going further, we quote 3$\sigma$ flux and luminosity limits in the 0.5-8 keV band.

Upper limits on X-ray SNe, and on X-ray observations in general, are notoriously unreliable. Firstly, different backgrounds result in different number of counts, which can lead to different count rates and upper limits. Furthermore, in order to convert the counts to a flux, say using Chandra PIMMS, a model is needed. Not being able to fit the data, an arbitrary model is used by necessity. But in a case where there are no prior objects detected in the same class, such as this one, there is no known model fit to the emission, and the model must be carefully chosen. Finally, in order to get an intrinsic flux and luminosity, a column density must be selected. Without further information being available, many authors use a Galactic column in the direction of the object. But this does not account for the presence of any material around the SN, or even in the intervening host galaxy, and can be a severe underestimate. Specifically for SNe, if there is material around the SN due to mass-loss from the progenitor or a companion star, this will add to the column density and lead to a higher X-ray luminosity. Neglecting this material means underestimating the column density as well as the SN luminosity, sometimes by an order of magnitude or higher.

The diversity of upper limit estimations is clearly seen in the literature. In their analysis of several Type Ia SNe observed (but not detected) with Chandra, \citet{hughesetal07} did a very careful estimation of the background. They considered various emission models for the flux, with different emitting temperatures, and several different intrinsic column densities in addition to the Galactic column towards the source. The X-ray flux varies in each case, and they provided a listing of the flux corresponding to the different options. Conversely, in their analysis of two Type Ia SNe, \citet{sandetal21} consider one background region, one model (a  power-law with defined power-law index of 2.0) and the Galactic column, and quote upper limits. This may be true if some justification was offered for using a  power-law model, and for  assuming that there is no additional material present around the SN, including any galactic column within the host galaxy, but such justification is not offered. The fact that it leads to a low luminosity is a circular argument, since the intrinsic luminosity is obtained by assuming a low column density. This may place some doubt on their claims that the data are among the most constraining for any Type Ia SNe beyond 10 Mpc. \citet{marguttietal14}  carry out a more careful estimation of the internal absorption column in the host galaxy of SN 2014J, and suggest that Inverse Compton emission will dominate in the first few days, typical of a fast shock in a low density medium. This is a likely and often used, if as yet unproven approximation, for a normal Type Ia SNe. \citet{ri12} use the column density towards the host galaxy, and assume a thermal bremsstrahlung spectrum with a 10 keV temperature in their analysis of Swift SNe, without justifying the model. They then proceed to derive upper limits for the mass lost by Type 1a SN systems via a possible companion star, notwithstanding the fact that by assuming a Galactic column towards the source, they have neglected the additional column density from any intrinsic material that was lost and surrounds the Type Ia SN.

Of the above, \citet{hughesetal07} is the only paper to have considered SNe of the Ia-CSM variety. Although  SN 2018fhw is somewhat different from SNe of Type Ia-CSM in that it has lower  H$\alpha$ luminosity, and is found in a low luminosity galaxy, the presence of the H$\alpha$ line suggests a high density medium around the SN (\S \ref{sec:disc}), perhaps at some distance. Given the high density, which will result in a slowing down of the shock wave, the most likely emission mechanism in these SNe would be thermal emission, due to thermal bremsstrahlung combined with line emission. Thermal emission depends on the square of the density of the emitting plasma, and therefore of the density of the surrounding medium, and will thus be higher at high densities. It was postulated as the X-ray emission for SN 2012ca by \citet{bocheneketal18}. We take this to be the preferred model for the emission, although we also quote a flux using a power-law model for comparion. Note that the presence of the high density will lead to a high column density around the SN, which must be taken into account in addition to the Galactic column towards the source, in order to calculate the intrinsic luminosity. This was done in \citet{hughesetal07}.

We account for intrinsic material surrounding the SN by assuming a range of larger intrinsic column densities, and quote values for the flux determined assuming this additional column along with the Galactic one. In the case of SN 2012ca, \citet{bocheneketal18} found that the column density was likely higher than 1 $\times 10^{22}$ cm$^{-2}$ cm$^{-2}$ for a temperature of a few keV. Not knowing the density and its variation with radius, herein we take into account intrinsic column densities ranging from 1 $\times 10^{21}$ cm$^{-2}$ cm$^{-2}$ to 2 $\times 10^{22}$ cm$^{-2}$ cm$^{-2}$. These account for any column in the host galaxy as well as any material surrounding the SN.  For each SN, we describe the  thermal emission using the PLASMA/APEC model in PIMMS, corresponding to thermal bremsstrahlung combined with line emission. The column density is given by the Galactic column towards the source, plus any additional column density given in the table, representative of material in the host galaxy and around the source. We also consider a range of X-ray temperatures from 1.53 keV to 9.67 keV.  Solar abundances are used, and 3$\sigma$ results quoted. For the power-law model, representative of Inverse Compton or synchrotron emission, we use a power-law index of 2 and the Galactic column. Finally we also give for comparison the upper limit obtained using the {\em srcflux} command in Sherpa, using the default power-law model with $\Gamma=2$ and the Galactic column. 

The Galactic column density towards SN 2018fhw, obtained using Colden\footnote{https://cxc.harvard.edu/toolkit/colden.jsp}, is 2.82 $\times 10^{20}$ cm$^{-2}$ \citep{dl90}. A redshift of 0.017, and distance to the source of 74.2 Mpc \citep{vallelyetal19} are adopted.  Our values for the flux and the resultant luminosity, calculated from the source counts using Chandra PIMMS\footnote{https://cxc.harvard.edu/toolkit/pimms.jsp}, are given in Table \ref{tab:flux18fhw}. We find 3$\sigma$ instrinsic  luminosities lying below about 3 $\times10^{39}$ erg s$^{-1}$.

\begin{table}
	\centering
	\caption{The unabsorbed flux and luminosity of SN 2018fhw, assuming various emission models with different parameters, and a distance of 74.2 Mpc. `Temp'=Temperature of the plasma. Column refers to the intrinsic column, in addition to the Galactic column of 2.82 $\times 10^{20}$ towards the source. 3$\sigma$ results are quoted.}
	\label{tab:flux18fhw}
	\begin{tabular}{lcccc} 
		\hline
		Model & Temp & Column & Flux & Luminosity\\
		            & (keV)   &  (cm$^{-2}$)          & (erg s$^{-1}$ cm$^{-2}) $ & (erg s$^{-1}$) \\
		\hline
		Power-law, $\Gamma$=2 &  &  & 1.6 $\times 10^{-15}$ & 1.05 $\times 10^{39}$ \\
		 (Using {\it srcflux})  &  &  &  1.6 $\times 10^{-15}$ & 1.05 $\times 10^{39}$ \\
		 \hline
		Plasma/APEC & 1.93  &  &  1.24 $\times 10^{-15}$ & 8.1 $\times 10^{38}$ \\
		Plasma/APEC & 4.85   &   &  1.54 $\times 10^{-15}$ & 1. $\times 10^{39}$ \\
		Plasma/APEC & 9.67  &   &  1.7 $\times 10^{-15}$ & 1.1 $\times 10^{39}$ \\
		\hline
		Plasma/APEC & 1.53  & 1. $\times 10^{21}$   &  1.3 $\times 10^{-15}$ & 8.5 $\times 10^{38}$ \\
		Plasma/APEC & 1.53  & 5. $\times 10^{21}$  & 1.8 $\times 10^{-15}$ & 1.2 $\times 10^{39}$ \\
		Plasma/APEC & 1.53  &  1. $\times 10^{22}$ & 2.45 $\times 10^{-15}$ & 1.6 $\times 10^{39}$ \\
		Plasma/APEC & 1.53  &  2. $\times 10^{22}$ & 3.95 $\times 10^{-15}$ & 2.6 $\times 10^{39}$ \\
		\hline
		Plasma/APEC & 3.06  & 1. $\times 10^{21}$   &  1.5 $\times 10^{-15}$ & 9.8 $\times 10^{38}$ \\
		Plasma/APEC & 3.06  & 5. $\times 10^{21}$  & 1.85 $\times 10^{-15}$ & 1.2 $\times 10^{39}$ \\
		Plasma/APEC & 3.06  &  1. $\times 10^{22}$ & 2.3 $\times 10^{-15}$ & 1.5 $\times 10^{39}$ \\
		Plasma/APEC & 3.06  & 2. $\times 10^{22}$  & 3.2 $\times 10^{-15}$ & 2.1 $\times 10^{39}$ \\
		\hline
		Plasma/APEC & 4.85  & 1. $\times 10^{21}$  & 1.63 $\times 10^{-15}$ & 1.1 $\times 10^{39}$ \\
		Plasma/APEC & 4.85  & 5. $\times 10^{21}$  & 2. $\times 10^{-15}$ & 1.3 $\times 10^{39}$ \\
		Plasma/APEC & 4.85  & 1. $\times 10^{22}$  & 2.4 $\times 10^{-15}$ & 1.6 $\times 10^{39}$ \\
		Plasma/APEC & 4.85  & 2. $\times 10^{22}$  & 3.15 $\times 10^{-15}$ & 2.1 $\times 10^{39}$ \\
		\hline
	\end{tabular}
\end{table}

\section{Comparison to other SNe} 
\label{sec:comp}
In order to compare the upper limits obtained for SN 2018fhw with other SNe, it is essential to select SNe with similar properties, which is an impossible task since no similar SNe have been detected, or even observed, in X-rays.  As illustrated in section \ref{sec:disc}, it is likely that these SNe are surrounded by a high density medium. Typical Type Ia SNe are not expected to have a high density medium around them. In the category of SNe that display an H$\alpha$ line in the optical spectrum, SN 2012ca happens to be the only one that has been detected in X-rays, and it belongs to the Type Ia-CSM category. Its X-ray properties have been summarized by \citet{bocheneketal18}. At 554 days after outburst it had a flux (0.5-7 keV) of 7.6$^{+389}_{- 5.1} \times 10^{-14}$ erg s$^{-1}$ cm$^{-2}$, and a corresponding luminosity of 5.81$^{+298}_{-3.92} \times 10^{40}$  erg s$^{-1}$. At the second epoch of 745 days, the 0.5-7 keV flux was 3.68$^{+104}_{- 3.05} \times 10^{-14}$ erg s$^{-1}$ cm$^{-2}$, with a corresponding luminosity of 2.82$^{+79.9}_{-2.34} \times 10^{40}$  erg s$^{-1}$

Two other Type Ia-CSM SNe, SN 2005gj and SN 2002ic, have been previously observed in X-rays, although no X-rays were detected. In the following we study these SNe and compute upper limits for them, using a  treatment similar to SN 2018fhw.

SN 2005gj (ObsID 7241, PI: Pooley) was studied earlier by \citet{prietoetal07} and \citet{hughesetal07}. As before, we use a  1" source region, which  includes 2 counts in the 0.5-8 keV region, matching \citet{prietoetal07}. We used as background an annulus region stretching from 1" to 5". This background region contains 17 counts in the 0.5-8 keV range, which would indicate 0.71 counts in the source region. Using the method of \citet{kbn91}, this results in a 3$\sigma$ upper limit of 9.24 counts, or a count rate of 1.86 $\times 10^{-4}$ counts s$^{-1}$ using the 49.54 ks exposure time. Using this count rate, a Galactic column towards the source of 6.87 $\times 10^{20}$ cm$^{-2}$ \citep{dl90}, and a redshift of 0.0618, we compute the upper limits. We take into account power law as well as thermal models with varying input temperatures and intrinsic column densities, similar to those for SN 2018fhw above. The results are listed in Table \ref{tab:flux05gj}. In order to compare with  \citet{prietoetal07} we also calculated 2$\sigma$ results (not shown), and find that our results are consistent with those of \citet{prietoetal07}. Since \citet{hughesetal07} compared their flux limits with those of \citet{prietoetal07} and found that they were comparable, our results would therefore also be agreeable with \citet{hughesetal07} where overlap exists.  A more elaborate comparison with \citet{hughesetal07} is somewhat difficult because their results are quoted in different wavebands. We calculated a few 3$\sigma$ flux results in their wavebands (not shown), and they were generally agreeable.

\begin{table}
	\centering
	\caption{The unabsorbed flux and luminosity of SN 2005gj, assuming various emission models with different parameters, and a distance of 266 Mpc. `Temp'=Temperature of the plasma. Column refers to the intrinsic column, in addition to the Galactic column of 6.87 $\times 10^{20}$ towards the source. 3$\sigma$ results are quoted. }
	\label{tab:flux05gj}
	\begin{tabular}{lcccc} 
		\hline
		Model & Temp & Column & Flux & Luminosity\\
		            & (keV)   &  (cm$^{-2}$)          & (erg s$^{-1}$ cm$^{-2}) $ & (erg s$^{-1}$) \\
		\hline
		Power-law, $\Gamma$=2 &  &  & 1.3 $\times 10^{-15}$ & 1.1 $\times 10^{40}$ \\
		 (Using {\it srcflux})  &  &  &  1.13 $\times 10^{-15}$ & 9.5 $\times 10^{39}$ \\
		 \hline
		Plasma/APEC & 1.93  &  &  9.78 $\times 10^{-16}$ & 8.2 $\times 10^{39}$ \\
		Plasma/APEC & 4.85   &   &  1.42 $\times 10^{-15}$ & 1.2 $\times 10^{40}$ \\
		Plasma/APEC & 9.67  &   &  1.63 $\times 10^{-15}$ & 1.4 $\times 10^{40}$ \\
		\hline
		Plasma/APEC & 1.53  & 1. $\times 10^{21}$   &  1. $\times 10^{-15}$ & 8.4 $\times 10^{39}$ \\
		Plasma/APEC & 1.53 & 5. $\times 10^{21}$  & 1.58 $\times 10^{-15}$ & 1.3 $\times 10^{40}$ \\
		Plasma/APEC & 1.53 &  1. $\times 10^{22}$ & 2.44 $\times 10^{-15}$ & 2.1 $\times 10^{40}$ \\
		Plasma/APEC & 1.53 &  2. $\times 10^{22}$ & 4.49 $\times 10^{-15}$ & 3.8 $\times 10^{40}$ \\
		\hline
		Plasma/APEC & 3.06  & 1. $\times 10^{21}$   & 1.34 $\times 10^{-15}$ & 1.1 $\times 10^{40}$ \\
		Plasma/APEC & 3.06  & 5. $\times 10^{21}$  & 1.91 $\times 10^{-15}$ & 1.6 $\times 10^{40}$ \\
		Plasma/APEC & 3.06  &  1. $\times 10^{22}$ & 2.6 $\times 10^{-15}$ & 2.2 $\times 10^{40}$ \\
		Plasma/APEC & 3.06  &  2. $\times 10^{22}$ & 3.92 $\times 10^{-15}$ & 3.3 $\times 10^{40}$ \\
		\hline
		Plasma/APEC & 4.85  & 1. $\times 10^{21}$  & 1.58 $\times 10^{-15}$ & 1.3 $\times 10^{40}$ \\
		Plasma/APEC & 4.85  & 5. $\times 10^{21}$  & 2.17 $\times 10^{-15}$ & 1.8 $\times 10^{40}$ \\
		Plasma/APEC & 4.85  & 1. $\times 10^{22}$  & 2.83 $\times 10^{-15}$ & 2.4 $\times 10^{40}$ \\
		Plasma/APEC & 4.85  & 2. $\times 10^{22}$  & 4.03 $\times 10^{-15}$ & 3.4 $\times 10^{40}$ \\
		\hline
	\end{tabular}
\end{table}

SN 2002ic (ObsID 4449, PI: Hughes) was also studied by \citet{hughesetal07}. A  1" region includes 3 counts in the 0.5-8 keV region. For the background region we use an annulus from 1" to 6", which contains 12 counts in the 0.5-8 keV range. This would result in 0.3428 counts in the source region.  Using the method of \citet{kbn91}, this gives a 1/ 2/ 3 $\sigma$ upper limit of 4.8/ 7.7/ 11.3  counts. The method of \citet{kbn91} returns a confidence interval with a non-zero lower limit for the 1$\sigma$ and 2$\sigma$ ranges, although the lower-limit is 0 for the 3$\sigma$ case. This suggests a detection with a low significance $< 3 \sigma$. Using the method of \citet{lm83}, we calculate the significance of the `detection' to be 1.56 $\sigma$, which would not be considered a conclusive detection. Therefore we continue to plot the values as an upper limit. This gives a count rate of 2.64/ 4.23/ 6.21 $\times 10^{-4}$ counts s$^{-1}$ for the 18.2 ks exposure. Using this count rate, a Galactic column towards the source of 5.93 $\times 10^{20}$ cm$^{-2}$ \citep{dl90}, and a redshift of 0.0666, we compute the upper limits as above, assuming power law as well as thermal models with varying input temperatures and intrinsic densities. Results for SN 2002ic are given in Table \ref{tab:flux02ic}.

\begin{table}
	\centering
	\caption{The unabsorbed flux and luminosity of SN 2002ic, assuming various emission models with different parameters, and a distance of 285 Mpc. `Temp'=Temperature of the plasma. Column refers to the intrinsic column, in addition to the Galactic column of 5.93 $\times 10^{20}$ towards the source. 3$\sigma$ results are quoted.}
	\label{tab:flux02ic}
	\begin{tabular}{lcccc} 
		\hline
		Model & Temp & Column & Flux & Luminosity\\
		            & (keV)   &  (cm$^{-2}$)          & (erg s$^{-1}$ cm$^{-2}) $ & (erg s$^{-1}$) \\
		\hline
		Power-law, $\Gamma$=2 &  &  & 3.6 $\times 10^{-15}$ & 3.5 $\times 10^{40}$ \\
		 (Using {\it srcflux})  &  &  &  3.65 $\times 10^{-15}$ & 3.5 $\times 10^{40}$ \\
		 \hline 
		Plasma/APEC & 1.93  &  &  2.76 $\times 10^{-15}$ & 2.7 $\times 10^{40}$ \\
		Plasma/APEC & 4.85   &   &  4.09 $\times 10^{-15}$ & 4. $\times 10^{40}$ \\
		Plasma/APEC & 9.67  &   &  4.83 $\times 10^{-15}$ & 4.7 $\times 10^{40}$ \\
		\hline
		Plasma/APEC & 1.53  & 1. $\times 10^{21}$   &  2.86 $\times 10^{-15}$ & 2.8 $\times 10^{40}$ \\
		Plasma/APEC & 1.53 & 5. $\times 10^{21}$  & 4.66 $\times 10^{-15}$ & 4.5 $\times 10^{40}$ \\
		Plasma/APEC & 1.53 &  1. $\times 10^{22}$ & 7.37 $\times 10^{-15}$ & 7.1 $\times 10^{40}$ \\
		Plasma/APEC & 1.53 &  2. $\times 10^{22}$ & 1.4 $\times 10^{-14}$ & 1.4 $\times 10^{41}$ \\
		\hline
		Plasma/APEC & 3.06  & 1. $\times 10^{21}$   &  3.92 $\times 10^{-15}$ & 3.8 $\times 10^{40}$ \\
		Plasma/APEC & 3.06  & 5. $\times 10^{21}$  & 5.86 $\times 10^{-15}$ & 5.7 $\times 10^{40}$ \\
		Plasma/APEC & 3.06  &  1. $\times 10^{22}$ & 8.18 $\times 10^{-15}$ & 7.9 $\times 10^{40}$ \\
		Plasma/APEC & 3.06 &  2. $\times 10^{22}$ & 1.27 $\times 10^{-14}$ & 1.2 $\times 10^{41}$ \\
		\hline
		Plasma/APEC & 4.85  & 1. $\times 10^{21}$  & 4.64 $\times 10^{-15}$ & 4.5 $\times 10^{40}$ \\
		Plasma/APEC & 4.85  & 5. $\times 10^{21}$  & 6.68 $\times 10^{-15}$ & 6.5 $\times 10^{40}$ \\
		Plasma/APEC & 4.85  & 1. $\times 10^{22}$  & 8.97 $\times 10^{-15}$ & 8.7 $\times 10^{40}$ \\
		Plasma/APEC & 4.85  & 2. $\times 10^{22}$  & 1.31 $\times 10^{-14}$ & 1.3 $\times 10^{41}$ \\
		\hline
	\end{tabular}
\end{table}

The upper limits on the luminosity for SN 2018fhw are lower than those of both SN 2005gj and SN 2002ic. The flux limits for SN 2018fhw and SN 2005gj are comparable, with those of SN 2002ic being somewhat higher. Interestingly, for all the parameters considered, the upper limits on the X-ray luminosity of SN 2018fhw are lower than the corresponding values for SN 2012ca, whereas those for SN 2005gj and 2002ic are comparable or somewhat higher. This may be because this SN is intrinsically dimmer than SN 2012ca, whereas the other two, being Type Ia-CSM SNe, are more comparable.  Alternatively, it may be because the observation of SN 2018fhw occurred almost a year after the 2nd observation of SN 2012ca, when the density had further decreased. Without having  knowledge of the density of the ambient medium, and its variation with radius, in both sources, the reason is difficult to determine. 

In the case of SN 2011fe, \citet{hughesetal11}  calculated a 3$\sigma$ upper limit $L_x = 3.8 \times 10^{36}$ erg s$^{-1}$ in the 0.5-8 keV band, 4 days after explosion, assuming a power-law model with spectral photon index of 2. In the case of SN 2014J, \citet{marguttietal14} used Chandra to compute a 3$\sigma$ upper limit to the X-ray luminosity of  $L_x = 8.7 \times 10^{36}$ erg s$^{-1}$ for a power-law in the 0.3-10 keV band at an age of 20.4 days. These limits were calculated very early on in the evolution compared to the SNe discussed above, and give a much lower maximum luminosity.  As with many limits in Type Ia SNe, they were calculated assuming that the emission mechanism is inverse Compton scattering of photons by the electrons accelerated at the fast shock propagating in the low density wind. We expect the X-ray emission to be higher for a shock propagating in a much higher density medium. It is not surprising then that the X-ray limits are higher for SN 2018fhw. \citet{ri12} studied a total of 53 SNe observed with the Swift XRT. None were detected, and they found 3$\sigma$ upper limits in the 0.2-10 keV band ranging from $L_x = 2.7 \times 10^{38}$ erg s$^{-1}$ to $L_x = 2.3 \times 10^{41}$ erg s$^{-1}$. Some of these limits are high due to contamination from the galactic nucleus, as mentioned. Overall they are not very constraining for typical Type Ia SNe. The authors predicted mass-loss rates for an underlying wind region, which were computed assuming an equation for thermal emission. However, if the emission is predominantly non-thermal, due to inverse Compton  emission as assumed in other cases mentioned above, then their assumptions, and therefore the mass-loss rates, may not be correct. 

\citet{hughesetal07} compared their observations to a model of the interaction of SN 2002ic with a dense circumstellar medium (CSM) by \citet{ccl04}. That model had overpredicted the X-ray intensity of the SN by a factor of 4 compared to the upper limits found by \citet{hughesetal07}. They resolved the discrepancy by invoking the mixing of cool dense ejecta fragments with the forward shock region, to increase the amount of absorption. The same scenario was also invoked by them for SN 2005gj. While the model is no doubt interesting, it is not clear that it could fit the data for SN 2012ca, given the lower velocity that was seen in SN 2012ca compared to that used in their models. Plus as noted, SN 2018fhw has major differences with the Type Ia-CSM SNe, in terms of their location and the lower optical luminosity. Therefore we do not deem it practical to apply the same model to SN 2018fhw that was used for a Type Ia-CSM SN.

Many estimates of X-ray upper limits have further computed the maximum density in case of a constant density medium, or a wind mass-loss rate in case of a  wind medium. These assume either that the density is constant, or in the case of a  wind that the mass-loss parameters (mass-loss rate and wind velocity) are constant. As we show in \S \ref{sec:disc}, neither of these appear to be applicable to SN 2018fhw. The density is unlikely to be constant over the entire 1053 days, otherwise the intensity of the X-ray emission would exceed the upper limits, and would be high enough to make it detectable. The constancy of the H$\alpha$ line intensity for the first 150 days makes it unlikely that the SN is evolving in a constant parameter wind, although this is investigated below. In the case of a two-component medium, with dense clumps in a low density wind, it is possibly that the lower-density medium is a constant parameter wind, but there is not sufficient data to constrain it. In any case it is unlikely that the X-ray emission from this lower-density component would be comparable to that from the higher density clumps.The most likely configuration is that the density had a certain profile, perhaps nearly constant, for about 150 days, given the constancy in the H$\alpha$ intensity, and then decreased after that, with insufficient data to measure the amount of the decrease.

\section{Discussion}
\label{sec:disc}
The H$\alpha$ line: The presence of the H$\alpha$ line suggests the existence of a dense medium around SN 2018fhw, required to provide the H$\alpha$ luminosity.  In order to evaluate the significance of the non-detection and upper limits on SN 2018fhw, and see how constraining they are, we estimate the density of the medium required to provide the observed H$\alpha$ luminosity, and the resulting intensity of the expected X-ray emission from the SN.  We expect that the X-ray emission mechanism will be thermal bremsstrahlung as outlined above, due to the high density medium.  The magnitude of the X-ray emission (proportional to the density squared) depends crucially on the density of the surrounding medium, and therefore on the mechanism that gives rise to the H$\alpha$, which is unclear. V19 found the FWHM of the H$\alpha$ line to be $\approx$ 1390 $\pm$ 200 km s$^{-1}$ up to 148d. One assumption is to consider this as a proxy for the shock velocity in the dense medium, although we do allow for other velocities. Since the typical shock velocity of a Type Ia SN is of order 10,000 km s$^{-1}$ \citep{wangetal09}, it is immediately apparent that a high density would be needed to reduce it to such a low value.

\subsection{Electron Scattering}
The electron scattering optical depth $\tau_{es}$ is the product of the scattering cross section, the electron density and the size of the region.  We can assume Thomson scattering, and use the Thomson scattering cross section $\sigma_e=6.65 \times 10^{-25}$ cm$^2$.   A velocity of 1400 km s$^{-1}$ for 150 d gives a region of size about 1.8 $\times 10^{15}$ cm.  The inferred density in the first 150 days is of order  10$^8$ cm$^{-3}$ (see below), therefore $\tau_{es} < 1$. Electron scattering would become important if the density or the radius were an order of magnitude larger. The electron density would need to increase to $8 \times 10^8$ cm$^{-3}$, for $\tau_{es} \approx 1$, which seems unlikely as it would have resulted in a much larger H$\alpha$ flux. The shock velocity could possibly be higher than the FWHM of the lines, but a velocity (and therefore radius) an order of magnitude larger would not be consistent with expansion in such a high density medium. At the time of Chandra observation the radius can grow to 10$^{16}$ cm, but as shown below the density $n_e < 5 \times 10^6$ cm$^{-3}$. Thus $\tau_{es} < 1$ again. We can therefore conclude that at the epochs considered in this study, electron scattering is not important.  \\

\citet{hc18} have studied the effect of electron scattering on lines emitted as a result of SN interaction with a dense circumstellar medium. In this case the line width would be at least partially due to electron scattering. The observational signature is broad wings on a narrow line feature. The line profile would be exponential, with the wings strengthening with optical depth. The shape of the lines  calculated by them does not appear to match those seen in SN 2018fhw. K19 maintain that the H$\alpha$ feature is well fit by a gaussian profile. V19 also fit the emission line with a gaussian. No narrow line feature is visible. \\

Given these considerations, we do not consider electron scattering as a viable explanation for the line broadening in our study. 

\subsection{Density of the Surrounding medium:} 

In SN 2012ca the high luminosity of the H$\alpha$ line, observed starting 50 days from the time of explosion, and the high Balmer decrement (3-20), was interpreted as being due to collisional excitation of the line. In SN 2018fhw, V19 place a lower limit of 2 on the Balmer decrement before 100 days, and a lower limit of 5 after this, consistent with the lower limit of 6 by K19. Given the Balmer decrement, it is less likely that the H$\alpha$ line is due to case B recombination. The constant flux was interpreted by V19 as evidence that the H$\alpha$ arises from CSM interaction rather than being swept-up material from a non-degenerate companion. In the following we therefore assume that the H$\alpha$ line arises due to  a high density CSM.\\

We estimate the density under various scenarios.

\begin{itemize}
\item{\em Case B recombination:} As mentioned above, this is unlikely, given the high value of the Balmer decrement, exceeding the nominal value of 3 \citep{osterbrock89}.  Given the uncertainties, and the fact that selective dust extinction could be a possibility, we still consider it here.  The Case B H$\alpha$ luminosity per unit volume is $L_{H\alpha} = n_e\,n_p\,\alpha_B h {\nu}_{\alpha}$, with the recombination coefficient $\alpha_B$ = 2.6 $\times 10^{-13}$ cm$^3$ s$^{-1}$ at a temperature of 10$^4$ K. The radius calculated above, 1.8 $\times 10^{15}$ cm, can be used to compute the volume. The emission at this stage should be arising from a thin shell (encased between the forward and reverse shocks) with a shell thickness of about 0.1 \citep{cf94}. The observed H$\alpha$ luminosity then gives a number density of 2.1 $\times 10^8$ cm$^{-3}$ up to at least 150 d.  Variations in the velocity would cause the density to change. An average velocity v$_{sh}$ km s$^{-1}$ would result in a particle density 2.1 $\times 10^8$ (v$_{sh}/1400)^{-3/2}$ cm$^{-3}$. Thus, even for a velocity of 5000 km s$^{-1}$, which would be at the higher end, the density would be $> 3 \times 10^7$ cm$^{-3}$.
\item{\em Radiative shocks:}  Depending on the density of the medium, the shock expanding into the medium can become radiative. The maximum luminosity one can get from a shock propagating in a region of radius $a$ and density $\rho$ is $2 \pi a^2 \rho v_{sh}^3$. This luminosity must be at least as large as the observed H$\alpha$ luminosity. The density of this region must therefore be  $\rho > L_{H\alpha} /(2 \pi a^2 v_{sh}^3$). The radius of the region $a$ should be smaller than the shock radius. If we take it to be 10\% of the radius, then we get a minimum density at least 2 $\times\, 10^8$  cm$^{-3}$.  The radius could be smaller if the emission was due to clumps or clouds in a low density medium, as in SN 2012ca. If the radius decreases the density would proportionately increase, so it is clear that a high density is needed to get the observed H$\alpha$ from a radiative shock. It can be shown easily that at this density, a 1400 km s$^{-1}$ shock will cool in about two weeks, so the assumption of a radiative shock is warranted.
 \item{\em Balmer Dominated Shocks:} These happen when a fast shock travels into a partially neutral medium. Given the high H$\alpha$  luminosity and the tentative low velocity of about 1400 km s$^{-1}$ for the shock, Balmer dominated shocks are unable to account for this high luminosity even assuming a very high neutral fraction close to 1 \citep{cr78}.
\item{\em Collisional excitation:} This was the reason given for the  high Balmer decrement in the case of SN 2012ca. For this mechanism to be effective,  densities $> 10^8$  cm$^{-3}$ are required \citep{du80}.
\item{\em Recombination following ionization by X-ray emission from  the shock wave:} The X-ray emission from the shock itself can ionize the surrounding medium, which on recombination gives rise to the H$\alpha$. \citet{cf94} showed that in this scenario the H$\alpha$ luminosity must be less than about 5\% of the X-ray luminosity. Thus the minimum X-ray luminosity must be at least 20 times the H$\alpha$ luminosity, or $> 4 \times 10^{39}$ erg s$^{-1}$, for the X-ray emission to give rise to the H$\alpha$. 
\end{itemize}

\subsection{X-ray luminosity} The above density computations all suggest a high X-ray luminosity {\em at the time the shock was expanding in the high density medium, or up to 150d}.  The luminosity can be written as L$_x = n_e^2 \,\Lambda$ V erg s$^{-1}$, where V is the volume and $\Lambda$ the cooling function. The primary coolant at these temperatures is thermal bremsstrahlung combined with line emission. For a density of $ 10^8$ particles cm$^{-3}$, and temperature a few times 10$^7$ K, the luminosity is L$_x = 10^{16} \times 3.5 \times 10^{-23}\, $V erg s$^{-1}$, assuming the density remains constant. The X-ray luminosity at around 150d would then be  $L_x \sim 2.6 \times 10^{39}$ erg s$^{-1}$.  Note that this would be close to the X-ray luminosity required to account for the observed H$\alpha$ luminosity due to X-ray ionization. In general, assuming other parameters don't charge, we can write the X-ray luminosity as $L_x \sim 2.6 \times 10^{39} n_8^2$ erg s$^{-1}$, where $n_8$ is the density in terms of 10$^8$ cm$^{-3}$. If the density were much lower than 10$^8$ cm$^{-3}$ it may be difficult to get the observed H$\alpha$ luminosity. The Chandra observation happened 903 days later in the SN frame. This means that the shock would have expanded much further. The simplest (unlikely) assumption is that the density and velocity remained constant. In that case the radius, and therefore the volume, would increase proportionally, and the X-ray luminosity would increase to $L_x \sim 9.1 \times 10^{41} n_8^2$ erg s$^{-1}$. For the nominal values, this would exceed the upper limits in Table \ref{tab:flux18fhw}.  If the velocity decreased with expansion, the radius increase would be somewhat lower, but still larger than the upper limits. Thus, it appears likely that the density and velocity did not remain constant, and certainly did not increase, which is consistent with expectations.

A common assumption made for core-collapse SNe is that they evolve in a wind with constant parameters (mass-loss rate and wind velocity) from the time of explosion. Although it is unlikely that this was the case for SN 2018fhw, it is illustrative to investigate the mass-loss rate needed to provide the luminosity in this scenario. The density in this case decreases as r$^{-2}$, and the X-ray emission can be shown to decrease as $L_x \sim t^{-1}$ \citep{flc96, dg12}.   

\citet{cf03} estimate the free-free X-ray luminosity from a SN shock:

\begin{equation}
L_x \approx 3. \times 10^{39}\;g_{ff}\;C_n\;\left(\frac{{\dot{M}}_{-5}}{v_{w1}}\right)^2\,t_{10}^{-1}   \;\; {\rm erg} \; {\rm s}^{-1}
\label{eq:lx}
\end{equation}

\noindent 
where $g_{ff}$ is the gaunt factor, of order unity, $C_n = 1$ for the forward shock, ${\dot{M}}_{-5}$ is the mass loss rate in units of 10$^{-5} M_{\odot}$ yr$^{-1}$, $v_{w1}$ is the wind velocity in units of 10 km s$^{-1}$, and $t_{10}$ is time in units of 10 days. This equation assumes electron-ion equilibration. It returns an upper limit to the mass-loss rate at the time of Chandra observation of $\approx 10^{-4} $ M$_{\odot}$ yr$^{-1}$ for a wind velocity of 10 km s$^{-1}$, which is not particularly constraining.

It is more likely that the density decreased after 150 days. Starting from the X-ray luminosity at 150 days as given above, and assuming it decreased as in a constant wind medium, i.e. $t^{-1}$, at the time of X-ray observations the X-ray luminosity would be $L_x \sim 3.7 \times 10^{38}$. This value is lower than the upper limits given in Table \ref{tab:flux18fhw}, thus showing that it is indeed viable. The density at this point would be around 2 $\times\; 10^6$ cm$^{-3}$, or perhaps somewhat lower, since we would expect the velocity to increase as the density decreases. Of course the value of the density could be arbitrarily lower than this, and the luminosity would be consequently lower. Not knowing the structure and density profile of the medium, it is difficult to make further conclusions. 

Given the calculations above for various density configurations, it can be seen that the X-ray upper limits for SN 2018fhw given in Table \ref{tab:flux18fhw} can accommodate a density of at most around 5 $\times\,10^6$ cm$^{-3}$ at the time of Chandra observations. This number is valid whether the density was constant throughout, or if it decreased after 150 days. However, if the medium had a constant density with this value throughout, it may be difficult to explain the H$\alpha$ luminosity.

As noted above, if the density early on is of order 10$^8$ cm$^{-3}$ and the shock velocity 1400 km s$^{-1}$ in this medium, then the shock would likely be radiative and the equation used above for radiative shocks would be needed. If the shock is radiative, it becomes difficult to account for both the H$\alpha$ and the X-ray luminosity via the radiative shock. These calculations depend on the temperature profile behind the radiative shock, and are beyond the scope of this paper. The cooling time for a radiative shock at this high temperature is inversely proportional to the density and directly to the velocity \citep{cf94}, so if the density were closer to 10$^7$ cm$^{-3}$ and the velocity increased somewhat, the cooling time would approach or exceed 5 months, and the shock would no longer be considered radiative in the period when the H$\alpha$ line was visible,

In the above we have assumed that the 1400 km s$^{-1}$ velocity seen is that of the SN forward shock into a spherically symmetric dense medium. Another possibility is a two-component medium, as postulated by \citet{bocheneketal18} for SN 2012ca, with higher density clumps embedded in a lower density surrounding medium. In this case the 1400 km s$^{-1}$ velocity would apply to the shock wave in the high density clumps, and the calculations above would similarly apply to the density in the clumps. The H$\alpha$ emission would arise from this medium. Since the clumps occupy a smaller fraction of the surroundings, in the Case B recombination scenario the density would have to be increased to give the same amount of H$\alpha$ emission. Specifically, if the clumps occupied a volume fraction $f$, the density would increase by $\sqrt(1/f)$ to compensate. In the case of a  radiative shock, if the size of each clump $a$ is much smaller, and there are $n$ clumps,  we could write $\rho > L_{H\alpha} /(2 \pi n a^2 v_{sh}^3$). The X-ray emission would be expected to primarily arise from the high density clump medium. If the density is higher, but the total volume is lower, the X-ray emission would be modified accordingly. The interclump medium would have a lower density, although at present there are not enough observations to evaluate this density. The shock in the interclump medium would be evolving faster than the shock wave within the clumps, but due to the lower density would not substantially contribute to the X-ray emission.

\section{Conclusions}
\label{sec:results}

To date, only one Type Ia SN, SN 2012ca, has been detected in X-rays. In this paper we used the ACIS-S instrument on the Chandra X-ray Observatory to study SN 2018fhw. No X-ray emission was detected at the position of the SN. Using the method of \citet{kbn91} we calculate 3$\sigma$ upper limits to the X-ray emission from the SN to not exceed 3 $\times 10^{39}$ erg s$^{-1}$.  Without additional information, it is difficult to place too many constraints on the properties of the medium around SN 2018fhw. The presence of the H$\alpha$ line suggests a density around 10$^8$ cm$^{-3}$ for the time that the H$\alpha$ emission was prevalent. At the time the X-ray observation was made, about 1053 days after explosion, the density can be constrained to lie below about 5 $\times\, 10^6$ cm$^{-3}$. Of course the density, and X-ray luminosity, could be much lower. 

The X-ray observation of SN 2002ic was made 275 days after explosion, or 260 days in the SN frame, while that of SN 2005gj was 77 days after explosion in the SN frame. SN 2018fhw was observed 1053 days after explosion. The Type Ia-CSM SN 2012ca was detected in X-rays at 554 days after explosion, with only a few counts about 200 days later. The X-ray emission depend on the density of the medium, and its distance from the SN, which probably varies from case to case. It is difficult to draw firm conclusions regarding the opportune time of observation that would increase the probability of detection in X-rays. It is possible that early in the evolution the column density may be too high, and thus all the X-rays could be absorbed, hampering a detection, as in 2002ic and 2005gj. For example, if the surrounding density was on order 10$^8$ cm$^{-3}$, and that region stretched for 10$^{15}$ or 10$^{16}$ cm, the column density would be 10$^{23}$ or 10$^{24}$ cm$^{-2}$. It is possible in this case that most, if not all of the emission could be absorbed. If the density was 10$^6$ cm$^{-3}$, the column density would decrease proportionately, but the X-ray emission, which goes as density squared, would decrease more. On the other hand, if the SN shock wave has crossed the high density region and is evolving in a lower density, the X-ray luminosity would be highly reduced. This may have happened in SN 2018fhw. It appears that SN 2012ca was caught at an appropriate time when the X-ray emission was still high, and while the column density was high, it allowed a significant detection. While it is tempting to suggest that the most opportune time to observe in X-rays may be when the H$\alpha$ line is visible, if the shock has not crossed a substantial portion of any  existing high density medium, the high column density could result in most of the X-ray emission being absorbed. Further observations at various wavelengths of Type Ia SNe with an H$\alpha$ line, and a determination of the density and density profile of the medium into which the SN is expanding, would certainly assist in their X-ray detection. 

\citet{szalaietal21} have studied interacting SNe using mid-IR data from the Spitzer Space Telescope. They observed the site of SN 2018fhw at two epochs, $\approx$ 250 and 450 days after explosion. They do not detect any source at that location. They do find a pointlike source about 2" away from the location of SN 2018fhw, which does not show any flux changes between the epochs. They conclude that it is not related to the SN, and that the SN remains undetected in the IR. These observations occurred at a much earlier epoch than the X-rays observations, but later than the period of H$\alpha$ line observations. The lack of an IR signal may suggest that the density was already lower at the time of the IR observations. It is interesting to note that several of the Type Ia-CSM mentioned in this paper have been detected in the mid-IR, including SN 2005gj, SN 2002ic, and SN 2012ca, as well as PTF11kx and SN 2013dn, as late as 1500 days after explosion. The non-detection of SN 2018fhw is consistent with the fact that its luminosity, even in the IR, would be lower than the 1a-CSMs. The latter seem to have a relatively homogeneous set of  lightcurves in the mid-IR.

\cite{dubayetal22} have studied the UV lightcurves of 1080 Type Ia SNe using archival data from  GALEX. They find that none of their data show convincing evidence of CSM interaction, and suggest that it is rare. This is not surprising, given that the number of Type Ia SNe showing an H$\alpha$ line, and thus having definite CSM interaction, are less than two dozen or so, compared to the thousands of Type Ia SNe that have been observed. It is of course possible that if the H$\alpha$ line is detectable over a small time period, it may have been missed in many SNe, especially low luminosity ones that were not followed with a high cadence for the first several hundred to thousand days.

It is difficult to overstate the significance of an X-ray detection of any of the class of Type Ia SNe with an H$\alpha$ line. A detection would:
\begin{itemize}
\item Firmly establish a new class of X-ray SNe. Only one such event has been detected in X-rays, and it  was an extreme Type Ia-CSM SN. Although it had a spectrum resembling that of a Ia, its luminosity was higher than average. SN 2018fhw, SN 2016jae, and SN 2018cqj/ATLAS18qtd are clearly sub-luminous Type Ia. An X-ray detection would firmly establish that Type Ia SNe are X-ray emitters.
\item Provide unambiguous evidence that at least a small number of Type Ia's are surrounded by, and possibly interacting with, high density H-rich material.
\item Provide an impetus to the single degenerate model for the formation of Type Ia's. Several recent observations, and especially the well-studied nearby event SN 2011fe, have seemed to favor the double degenerate model \citep{maoz14}.
\end{itemize}

 It is therefore imperative that we keep following up similar Type Ia SNe, such as SN 2016jae and SN 2018cqj, in X-rays. While the detection of any Type Ia in X-rays would be important, it is likely that the ones with a H line in their optical spectrum will have a high X-ray luminosity, and be detectable with current X-ray telescopes. It is also important to follow them at other wavelengths, especially radio, the detection of which would also point to CSM interaction. The eventual goal would be to determine the density profile of the medium in which the SN is expanding, thus providing some idea of the progenitor or companion mass-loss properties, and ideally identifying the progenitor itself. It is clear that the rewards of finding one are significant, and far outweigh the observational  risk of a non-detection.

\section*{Acknowledgements}
We thank the referee for a careful reading of the paper, and for their comments and suggestions. We are grateful to Roger Chevalier for comments. Support for this work was provided by the National Aeronautics and Space Administration through Chandra Award Number GO0-21058X issued by the Chandra X-ray Center, which is operated by the Smithsonian Astrophysical Observatory for and on behalf of the National Aeronautics Space Administration under contract NAS8-03060, and by National Science Foundation grant 1911061 awarded to the University of Chicago (PI: Vikram Dwarkadas). The scientific results reported in this article are based to a significant degree on observations made by the Chandra X-ray Observatory, including ObsIDs 4449, 7241, and 22445. This research has made use of software provided by the Chandra X-ray Center (CXC) in the application packages CIAO and Sherpa. It has also made use of the Chandra PIMMS and Colden packages. 

\section*{Data Availability}

All data are included in the tables in the article.



\bibliographystyle{mnras}
\bibliography{halpha} 




\bsp	
\label{lastpage}
\end{document}